\documentclass[aps,pra, superscriptaddress, twocolumn, letter]{revtex4-1}

\usepackage{b26_style}

\begin{document}

\title{Single-Spin Magnetomechanics with Levitated Micromagnets}
\author{J.~Gieseler}
\affiliation{Physics Department, Harvard University, Cambridge, MA 02318, USA.}
\author{A.~Kabcenell}
\affiliation{Physics Department, Harvard University, Cambridge, MA 02318, USA.}
\author{E.~Rosenfeld}
\affiliation{Physics Department, Harvard University, Cambridge, MA 02318, USA.}
\author{J.~D.~Schaefer}
\affiliation{Physics Department, Harvard University, Cambridge, MA 02318, USA.}
\author{A.~Safira}
\affiliation{Physics Department, Harvard University, Cambridge, MA 02318, USA.}
\author{M.~J.~A.~Schuetz}
\affiliation{Physics Department, Harvard University, Cambridge, MA 02318, USA.}
\author{C.~Gonzalez-Ballestero}
\affiliation{Institute for Quantum Optics and Quantum Information of the
Austrian Academy of sciences, A-6020 Innsbruck, Austria.}
\affiliation{Institute for Theoretical Physics, University of Innsbruck, A-6020 Innsbruck, Austria.}
\author{C.~C.~Rusconi}
\affiliation{Max-Planck-Institut f\"ur Quantenoptik, Hans-Kopfermann-Strasse 1, 85748 Garching, Germany.}
\author{O.~Romero-Isart}
\affiliation{Institute for Quantum Optics and Quantum Information of the
Austrian Academy of sciences, A-6020 Innsbruck, Austria.}
\affiliation{Institute for Theoretical Physics, University of Innsbruck, A-6020 Innsbruck, Austria.}
\author{M.~D.~Lukin}
\affiliation{Physics Department, Harvard University, Cambridge, MA 02318, USA.}

\date{\today}

\begin{abstract} 
We demonstrate a new mechanical transduction platform for individual spin qubits. In our approach,
 single micro-magnets are trapped using a type-II superconductor in proximity of spin qubits, enabling direct magnetic coupling between the two systems. 
Controlling the distance between the magnet and the superconductor during cooldown, we demonstrate three dimensional trapping  with quality factors around one million and kHz trapping frequencies. We further exploit the large magnetic moment to mass ratio of this mechanical oscillator to couple its motion to the spin degree of freedom of an individual nitrogen vacancy center in diamond. Our approach provides a new path towards interfacing individual spin qubits with mechanical motion for testing quantum mechanics with mesoscopic objects, realization of quantum networks, and ultra-sensitive metrology. 
\end{abstract}

\maketitle

Realizing coherent coupling between individual spin degrees of freedom and massive mechanical modes is an outstanding challenge in quantum science and engineering. Such a coupling could be used to create mechanical systems with a strong quantum non-linearity, allowing preparation of macroscopic quantum states of motion \cite{Rabl:2010cm}. In addition, mechanical systems can be used to mediate effective spin-spin interactions between distant spin-qubits \cite{Schuetz:2017bl}, enabling  applications  ranging from quantum information processing \cite{Rabl:2010kk} and sensing \cite{Kolkowitz:2012iw, Bennett:2013ga, Barson:2017ba} to tests of fundamental physics \cite{vanWezel:2011fy, Marshall:2003dj}.
One particularly promising approach is to engineer a strong spin-mechanical coupling via magnetic field gradients \cite{Rabl:2009fz, Arcizet:2011cg, Pigeau:2015kl, Mintert:2001dy, Hong:2012hv}. 
Achieving strong spin-resonator coupling requires a combination of high quality mechanical resonators, strong magnetic field gradients, and spin qubits with very long spin coherence times. 

\begin{figure}[!htb]
\begin{center}	
  \includegraphics[width=\columnwidth]{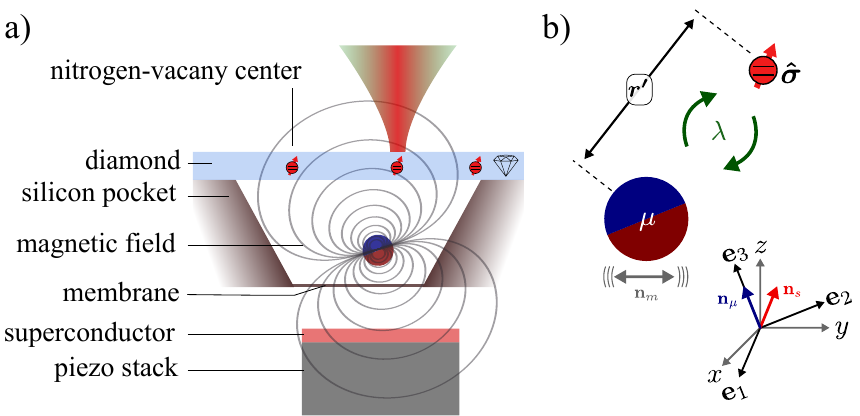}
  \caption{\label{fig:GeneralIdea}
Concept of the experiment.
 (a) Individual microscopic hard ferromagnets are isolated in microfabricated pockets and levitated with a type II superconductor, exploiting its flux trapping properties.
 The magnet's stray field allows for efficient coupling between the magnet's motional degrees of freedom and bulk nitrogen vacancy centers in the nearby diamond, which generally feature long spin coherence times.
 (b)
  The coupling $\lambda$ depends on the relative orientation of the NV-center, $\mathbf{n}_s$, the magnetic moment $\mathbf{n}_\mu$, the direction of motion of the magnet $\mathbf{n}_m$, the distance between the magnet and the NV-center $r'$, the magnet radius $a$ and the frequencies of the motion.
}
\end{center}
\end{figure}

In this Letter, we propose and demonstrate a new platform for strong spin-mechanical coupling based on levitated microscopic magnets coupled to the electronic ground state manifold of a single nitrogen vacancy (NV) center in diamond (\figref{fig:GeneralIdea}). 
The key idea is to utilize a levitated magnet that is 
localized in free space by electromagnetic fields. In such a system, dissipation is minimized since there is no direct contact with the environment. Specifically, we make use of a levitating mechanical resonator based on magnetostatic fields. This approach  not only avoids clamping losses, but also circumvents photon recoil and heating associated with optical levitation  \cite{RomeroIsart:2010iv, Chang:2010jn, Jain:2016fj} and is therefore predicted to yield large mechanical quality factors \cite{RomeroIsart:2012hg, Cirio:2012gka}.

In addition, the levitated magnet naturally generates the strong magnetic field gradient that is required for spin-mechanical coupling. We specifically demonstrate the coupling to an individual NV-center, one of the most studied color centers in diamond \cite{Doherty:2013bu}. Besides optical initialization and readout, the NV-center features long coherence times even at room temperature, which makes it an attractive candidate for scalable quantum networks in the solid state \cite{Rabl:2010kk}, quantum sensing \cite{Taylor:2008cp, Maze:2008cs, Kolkowitz:2015fx} and quantum communication \cite{Hensen:2015dw}.

Before proceeding, we note that low dissipation mechanical resonators based on magnetic levitation have been explored previously \cite{Barowski:1993uya, Gloos:1990ta}.
However, experiments with superconducting levitation have so far been limited to millimeter-scale magnets \cite{Druge:2014do, Barowski:1993uya, Nemoshkalenko:1990ku, Hull:1999jb}.
A recent experiment \cite{Tao:2019kl} demonstrated superconducting levitation of micro-magnets, but without spin-mechanical coupling and with much lower frequencies and Q-factors than shown here.
Levitated spin-mechanical systems in which the spin is hosted inside the resonator have been implemented with nano-diamonds containing NV-center defects trapped in optical tweezers \cite{Neukirch:2013ua, Hoang:2016ky}, Paul traps \cite{Kuhlicke:2014dm, Delord:2017cb, Alda:2016ce} and magneto-gravitational traps \cite{Hsu:2016cs,OBrien:2019kc}.
Nonetheless, the challenge remains to integrate these systems with strong magnetic field gradients, long coherence NV-centers and operation under ultra-high vacuum conditions.
Our approach fulfills all these criteria simultaneously (\figref{fig:GeneralIdea}).

\begin{figure}[!tb]
	\includegraphics[width=\columnwidth]{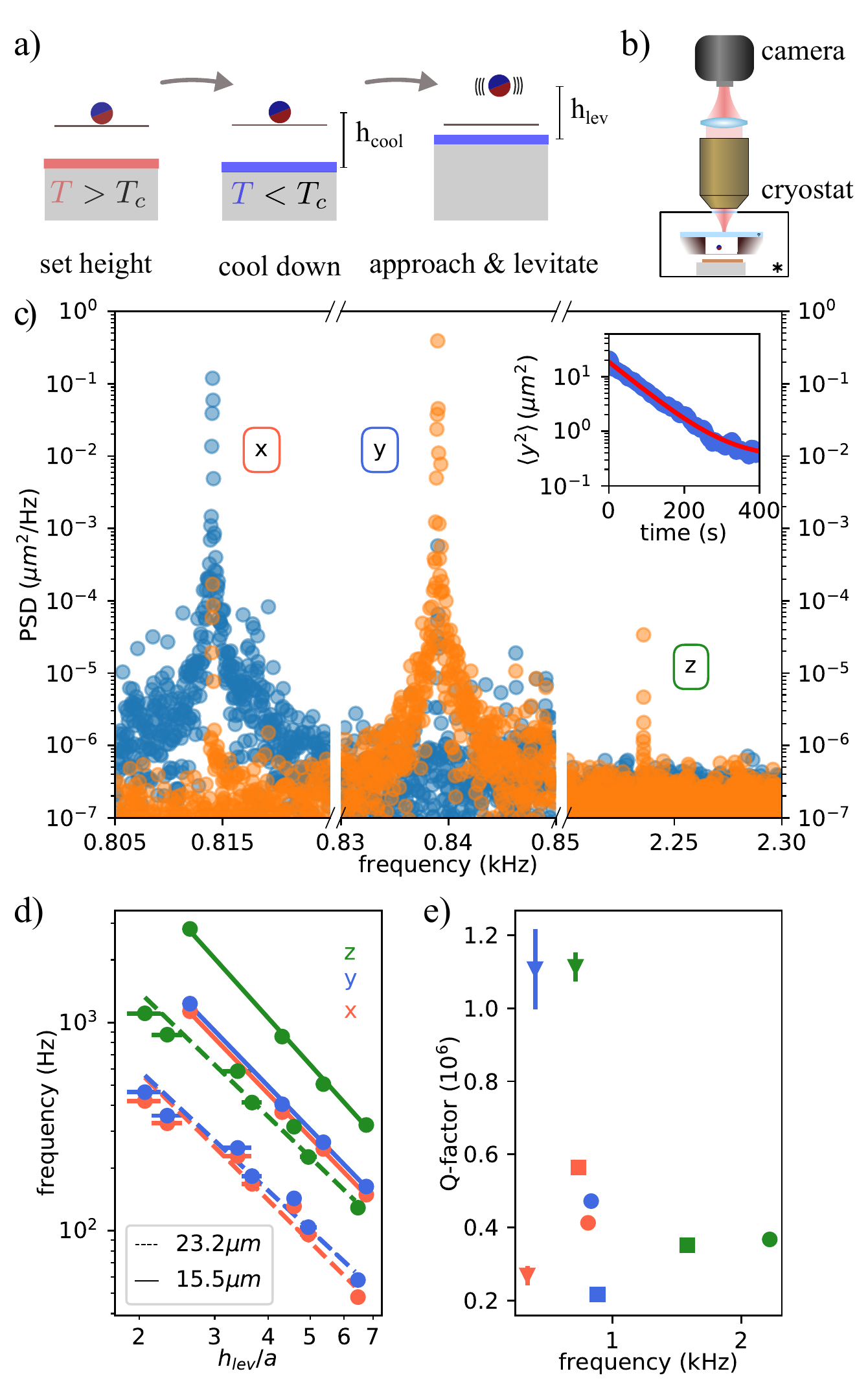}
	\caption{Mechanical properties of levitated micro-magnets.
      (a) To levitate the magnets, we adjust the magnet-sc distance above the critical temperature $T_c$. Then we cool the sc below $T_c$ to freeze in the magnetic flux from the magnet. After cooldown, we reduce the distance until the magnet begins to levitate.
    (b) We observe the magnet motion through a microscope objective and record its motion with a video camera.
	(c) Power spectral density extracted from video analysis of the magnet motion in the vertical (horizontal) direction, corresponding to the data shown as circles in (e).
	The inset shows a typical ringdown measurement of the y-mode ($\omega_y = 0.839 \kHz$) from which we extract the Q-factor.
	(d) Frequencies of center-of-mass motion as a function of the levitation height normalized to magnet radius. Dashed (solid) lines show the three center-of-mass modes for magnet 1 (2).
	(e) Q-factor as a function of trap frequency for magnet 2. Each symbol corresponds to a different levitation experiment and the different colors corresponds to the three different center-of-mass modes.
	}
	\label{fig:Motion} 
\end{figure}

\paragraph*{Levitation of micromagnets}
We levitate single hard micro-magnets with a thin film of the type-II superconductor (sc), yttrium barium copper oxide (YBCO)  (c.f. ~\figref{fig:GeneralIdea}).
Since we do not apply additional magnetic fields, the micromagnet is the only magnetic field source.
Thus, the magnetic flux through the YBCO film is determined only by the distance $\hcool$ between the magnet and the YBCO film, the orientation of the magnet $\thetacool$, the magnetization of the magnet, and its radius $a$.
After cooling the YBCO film below its critical temperature $\Tc \approx 90\rm K$, it becomes superconducting and magnetic flux that penetrates the film is frozen in. 
As a consequence, below $\Tc$, motion of the magnet induces currents in the superconductor that counteract changes in the magnetic field.
This allows for stable 3D trapping using a procedure illustrated in (\figref{fig:Motion}a).
The levitation height  ($\hlev$) and trapping frequencies $\omega_j$ $(j = x, y, z)$ depend on the conditions during cooldown, which we can control by adjusting the relative distance between the superconductor and the particle during cooldown ($\hcool$).

We observe the levitated magnet through a long working distance microscope objective that is positioned outside the vacuum chamber, and record its motion with a fast camera (\figref{fig:Motion}b).
From the video frames, we extract time-traces of the particle's center-of-mass position in the camera coordinate system ($x_c(t)$, $y_c(t)$) and calculate their power spectral densities (PSD).
The distinct peaks in the PSD shown correspond to the three center-of-mass modes $\w_j / 2\pi$ (\figref{fig:Motion}c).

\figref{fig:Motion}d displays the center-of-mass frequencies as a function of the normalized levitation height $\hlevn = \hlev / a$ for two particles with radius $\a_1=23.2\pm 0.7\um$ and $\a_2=15.5\pm 0.3\um$, respectively. The lines are a fit to a power law $f(\hlevn) = \levf \hlevn^{-\levexp}$, where $\levf$ is the frequency in the limit when the gap between the particle and the superconductor goes to zero. In our experiment, we find $\levf= (2.3\pm0.4, 2.4\pm0.4, 5.6\pm1.0)\kHz$ and $\levexp = (1.9\pm0.1, 2.1\pm0.1, 2.0\pm0.1)$ for particle 1 and $\levf= (8.8\pm1.1, 9.5\pm 1.1, 25.2\pm3.3)\kHz$ and $\levexp = (2.1\pm0.1, 2.1\pm 0.1, 2.3\pm0.1)$ for particle 2, which is in good agreement with the expected value of $\levexp = 2.5$ from a simple dipole model [SI].
The measured center-of-mass frequencies are comparable to those achieved with optical levitation \cite{Gieseler:2012bi} and significantly exceed motional frequencies in Paul traps \cite{Huillery:2019vz} and magneto-gravitational traps \cite{Hsu:2016cs, OBrien:2019kc}.
However, the observed dependence of the maximum frequency on the particle radius is stronger than the dipole model's prediction of $\levf^{dp}\propto 1 / a$. We attribute this to the multi-domain nature of our particles and note that spherical particles as large as $a \sim 1 \um$, which we expect to be achievable with this technique, can be single domain \cite{Skomski:2003by}.

\figref{fig:Motion}e shows the Q-factors of magnet 1 for three different levitation heights for all three translational modes.
We measure the dissipation with ring down measurements, exciting one mode with an AC magnetic field and observing its energy decay (\figref{fig:Motion}c inset). 
From an exponential fit of the energy decay we extract the decay time $ 1/\damp_j$ and the Q-factor $Q_j=\w_j/\damp_j$ for each mode.
The measured Q-factors are around one million and depend only weakly on the trapping frequencies and thus on the levitation height.
Air damping can be ruled out at our experimental conditions with pressures below $10^{-5}~\mBar$ and the most likely source of dissipation is the magnet-superconductor interaction.
Note that, even though this Q-factor is somewhat lower than what has been demonstrated with non-magnetic optically levitated \cite{Jain:2016fj} and nano-fabricated mechanical resonators \cite{Norte:2016eq,Reinhardt:2016ef, Tsaturyan:2017kd, Ghadimi:2018kb}, it represents the state of the art for magnetized resonators \cite{Longenecker:2012ip, Fischer:2019fy} and the ultimate limit, in particular for magnets with $a<\mu$m, is still an open question.

\begin{figure*}[hbt]
	\includegraphics[width=\textwidth]{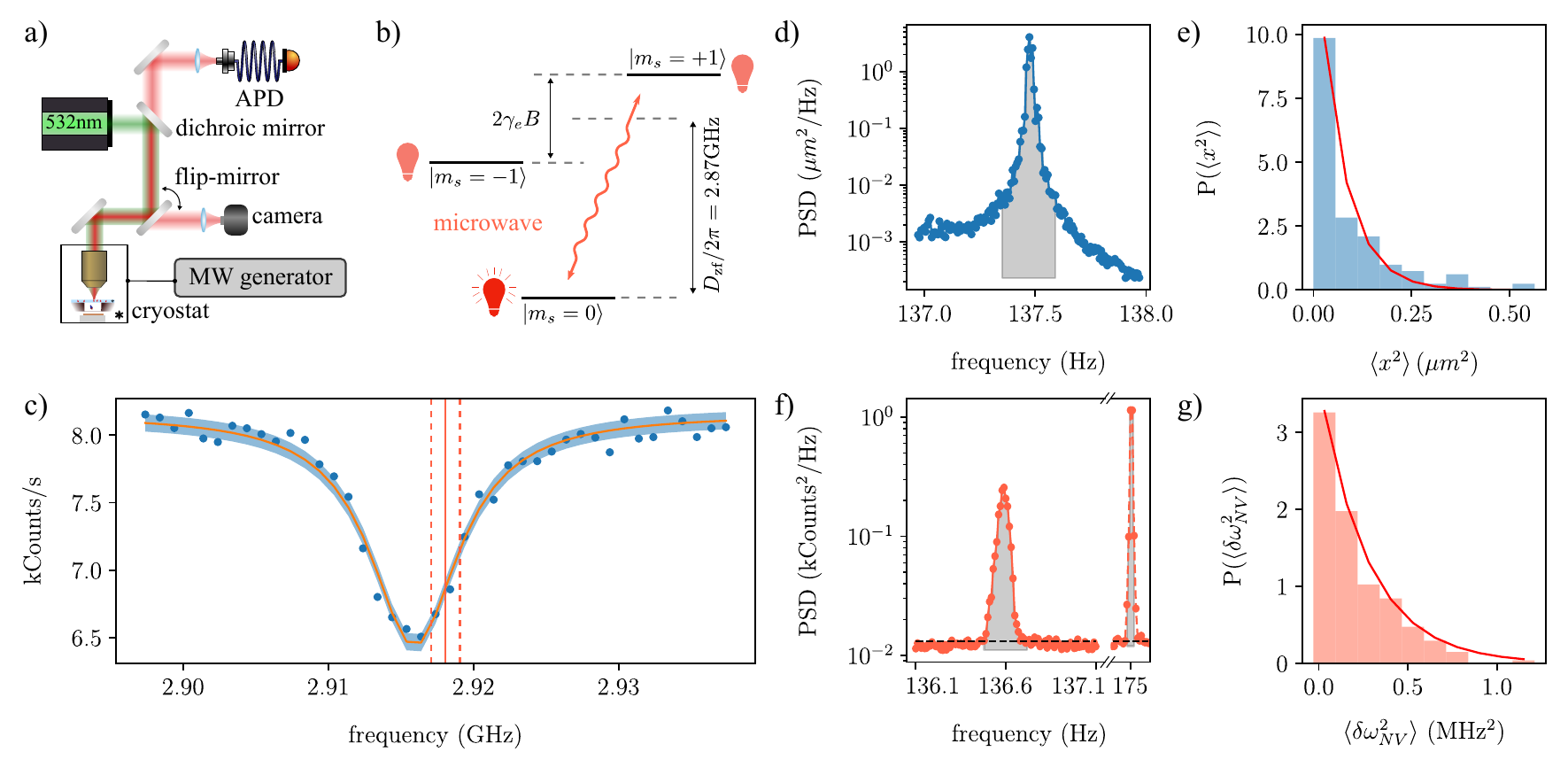}
	\caption{
Measurement of spin-motional coupling.
  (a) Schematic of the experiment. 
  (b) Nitrogen-vacancy ground state spin levels. Resonant microwaves drive transitions between the $m_s=0$ and $m_s=\pm1$ states. The transition frequency depends on the magnetic field, which depends on the magnet-NV distance. The $m_s=0$ is brighter than the $m_s=\pm1$ states. Hence, for near-resonant MW driving, the fluorescence intensity depends on the magnet position.
  (c) ODMR spectrum. The magnet motion is measured by recording a timetrace of the fluorescence photons while applying a MW tone at the steepest slope of the transition (2.918 GHz) (solid red line). We measure the slope by frequency modulating the MW tone (red dashed lines).
  (d) Power spectral density of the video timetrace and (f) fluorescence timetrace. The gray area corresponds to the numerical integration of the variance $\langle x^2 \rangle$, $\langle c_\text{NV}^2 \rangle$ and $\langle c_\text{cal}^2 \rangle$.
  The second sharp peak is the calibration peak due to the MW frequency modulation. The black dashed line represents the photon shot noise.
  Histograms of the variances reveal the thermal character of the motional state during both the (e) camera and (g) NV measurement which we fit (red line) to extract the  variances $\langle x^2 \rangle$ and $\langle \delta\omega_\text{NV}^2 \rangle$, respectively.
	}
	\label{fig:NV-data} 
\end{figure*}

\paragraph*{Coupling to NV-center}
Next, we demonstrate coupling the motion of a levitated micromagnet to the electronic spin associated with an individual negatively charged NV-center.
In our sample, NV-centers are hosted inside a diamond slab and implanted $d_\text{impl}\sim 40 \,{\rm nm}$ below the diamond surface. The diamond replaces the glass slide of the previous experiment and is placed across the pocket that contains the magnet. The pocket is $\sim 80 \um $ deep and the magnet radius is $a_3 = 15.1\pm0.1 \um$.
We levitate the magnet with the superconductor $z_{md} = 44 \pm 5 \um $ below the diamond using the same method as before.
The NV-center is located at $(x_d, y_d) = (83, 29)\pm 5 \um$ with respect to the magnet center such that the distance between the center of the magnet and the NV-center is $|\mathbf{r}'| = \sqrt{(z_{md}+a_3)^2+x_d^2+y_d^2} = 99 \pm 5 \um$.
The NV-center's electronic ground state has spin $S = 1$ with the lower-energy $\ket{m_s=0}$ level separated from the $\ket{m_s=\pm1}$ levels by a zero-field splitting
$D_\text{zf}/(2\pi) \approx 2.87\, \text{GHz}$ and its symmetry axes $\mathbf{n}_\text{s}$ aligned along one of four crystallographic orientations set by the tetrahedral symmetry of the diamond lattice.
A microwave (MW) signal at the transition frequency $\omega_\text{MW}$, drives the transition $\ket{m_s=0}\to \ket{m_s=\pm1}$ which results in a decrease of the PL signal.
The spin-dependent PL of the NV defect is due to a non-radiative intersystem crossing decay pathway, which also allows for efficient spin-polarization in the $\ket{m_s=0}$ spin sublevel through optical pumping \cite{Tetienne:2012fv}.
The magnetic field dependent PL can therefore be used to optically detect magnetic fields \cite{Rondin:2014kd}, which we will use to sense the motion of the magnet (Fig.~\ref{fig:NV-data}b). 

Fig.~\ref{fig:NV-data}c shows the optically detected magnetic resonance (ODMR) spectrum of the NV-center with a fit to a Lorentzian corresponding to the $\ket{+1}$ transition.
The spectrum is measured with a home-built fluorescence microscope that we integrated with the cryostat, replacing the long working distance objective outside the chamber with a high NA objective inside the chamber to maximize the collection efficiency of the fluorescence photons (Fig.~\ref{fig:NV-data}a).
In the presence of a microwave tone, the spin mechanical coupling $\lambda_g$  causes a variation in PL, since a displacement $x$ of the magnet shifts the electron spin resonance by $\delta\omega_\text{NV} = (\lambda_g/x_\text{xp})x$, where $x_\text{zp} = \sqrt{\hbar \left/ 2 m \omega_x\right.}$ is the zero point motion, in this case of $24\pm1$fm.
To measure the magnet's motion with the NV, we excite one of its modes with a broadband fluctuating magnetic field, which drives it into a quasi-thermal state and allows us to observe it as a peak in the power spectral density of the NV PL counts (\figref{fig:NV-data}f).
We confirm that the peak is due to the moving magnet with the camera (\figref{fig:NV-data}d).
The camera and NV measurements are taken sequentially.
Notably, we observe a small systematic frequency shift of $\approx 1$ Hz between the NV and camera measurements,
which is likely due to the laser field turned on during the NV measurement.

\begin{figure}[hbt]
	\includegraphics[width=\columnwidth]{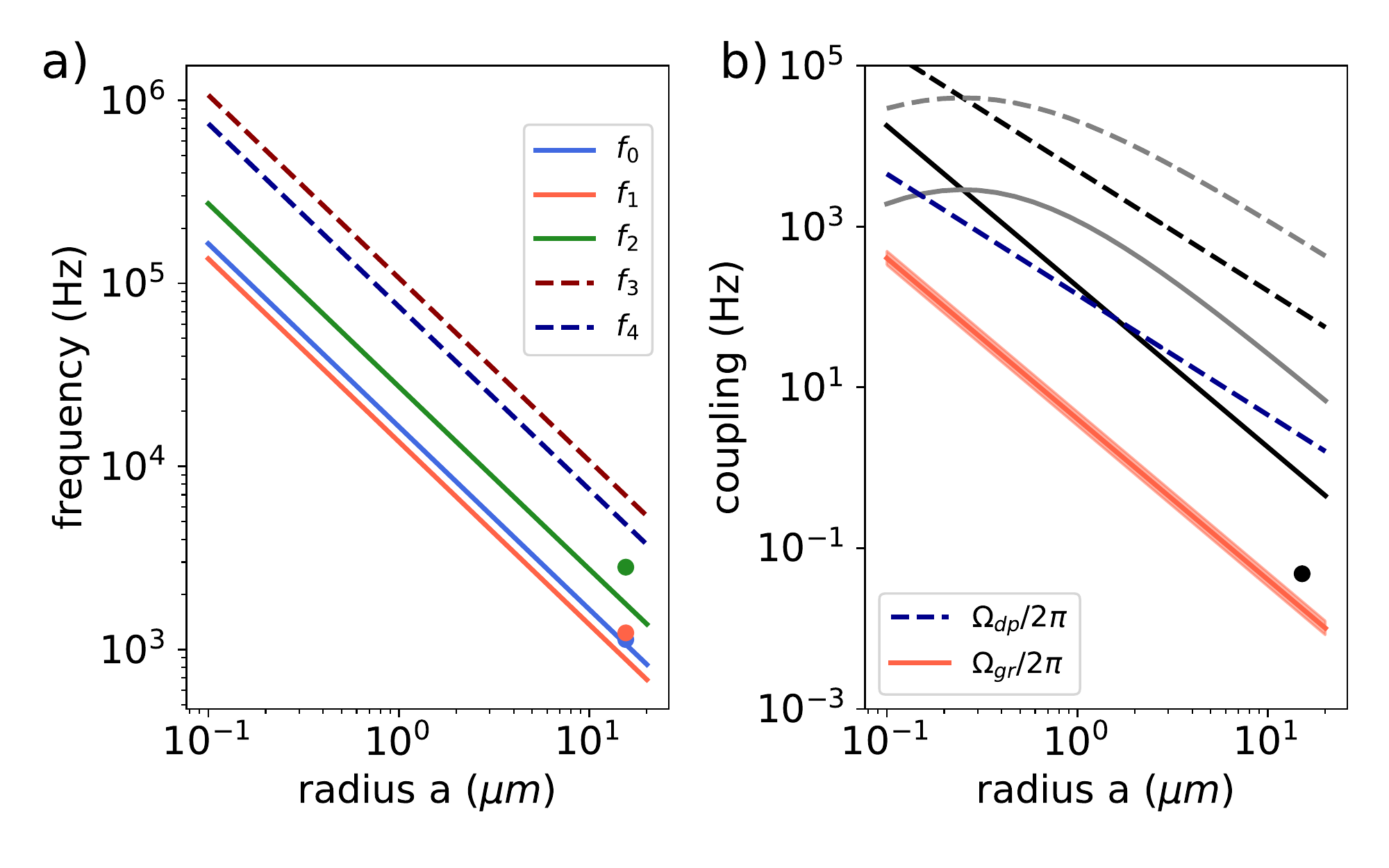}
	\caption{Prospect for frequencies and couplings. 
	(a) Mode frequency following dipole model \cite{Kordyuk:1998ht} for $\hlev/a=3$. The lowest (solid) lying modes correspond to the center-of-mass and the highest (dashed) lying modes to rotational motion. Data points are the experimental frequencies for particle 1 \figref{fig:Motion}d. MHz frequencies are predicted for sub-$\um$ particles.
	(b) Gradient (solid) and dipole-dipole (dashed) couplings, respectively. Straight lines show couplings for constant gap-particle size ratio $\bar{d}=d/a=5.5$ (color) and $\bar{d}=1$ (black). Curved lines (gray) show coupling for constant gap $d=250 \nm$ with a maximum coupling at $a=d$.
	The black data point is the experimental gradient coupling for particle 3.
	}
	\label{fig:Couplings} 
\end{figure}

The mean spectral power in the NV peak is $\langle c_\text{NV}^2 \rangle= s^2 \langle \delta\omega_\text{NV}^2\rangle$, where $s$ is the slope of the ODMR signal at the microwave frequency, which we measure by applying a calibration tone to the microwave signal. The mean spectral power in the camera measurement, $\langle x^2\rangle$, allows us then to extract the spin-mechanical coupling as $\lambda_g = x_\text{zp} \sqrt{\langle \delta\omega_\text{NV}^2\rangle \left/\langle x^2\rangle\right.}$. To measure the coupling and confirm the thermal character of the driven mode, we consider the area under the PSD integrated over a time interval much shorter than $1/\gamma$, and construct its distribution over repeated measurements.
For both the camera and the NV measurements, the distribution agrees with
an exponential distribution
$P(E)= \beta  \exp(-\beta E)$, where the decay constant $\beta$ is the inverse of the variances $\langle x^2 \rangle$ and $\langle \delta\omega_\text{NV} \rangle$ (\figref{fig:NV-data}e,g).
The resulting coupling strength is $48\pm 2 \mHz$, in satisfactory agreement with the theoretical value for the gradient coupling to a dipole
  $ \lambda_g =  \gamma_e \frac{\mu_0 \densitymag a^3}{r'^4} \, x_\text{zp} f_g(\theta)=2\pi \times (18\pm 3) \mHz f_g(\theta)$ (\figref{fig:Couplings}b).
  Here $f_g(\theta)$ is a function on the order of 1 that depends on the relative position and orientation of the NV-center and the magnet.

\paragraph*{Discussion}
We now discuss the prospects of using this system to achieve strong coupling. The minimal separation $d_q^\text{min}=|\r'|-\a$ between magnet and NV-center is given by the NV implantation depth and the onset of strong attractive surface forces that will make the magnet stick to the diamond surface.
For a given separation and assuming that the frequency scales as $\omega_j \left/2\pi\right.= \alpha a^{-n}$, the radius $a=(n+3) / (5-n) d_q^\text{min}$ yields the maximum gradient coupling for a dipolar particle.
A conservative gap $d_q^\text{min}=0.25\um$, $\alpha=15\kHz \, \mu m$, corresponding to our observations in Fig.\ref{fig:Motion}, and $n=1$ for the dipole model,
results in
$a=0.25\um$ and 
  $ \lambda_g\left/2\pi\right. \sim 2.6\kHz$.
Since the motional frequency can be reduced by adjusting the levitation height, one can even reach the elusive ultra-strong coupling regime $\lambda_g > \omega_j$ \cite{FornDiaz:2019br}.

The cooperativity $C = \lambda^2 Q \tilde{T}_2 \hbar \left/\left(2\pi k_B T\right)\right. > 1$ marks the onset of coherent quantum effects in a coupled spin-phonon system.
With a mechanical Q-factor of  $10^8$, which has been demonstrated in levitated systems \cite{Gieseler:2013ip}, the coupling exceeds the thermal decoherence rate $\Gamma_\text{th} / 2\pi = k_B T / (2\pi \hbar Q) = 0.8\rm kHz$ at $T=4\rm K$. 
NV-centers in bulk diamond, such as the sample used in our experiment, can have up to second long extended coherence times $\tilde{T}_2$ at these temperatures using spin manipulation such as multi-pulse dynamical decoupling sequences limited only by pulse errors \cite{Abobeih:2018bc, BarGill:2013dq}. The minimum spin manipulation (and therefore sensing) frequency in such sequences is set by the power spectrum of the noise. It is typically a few kHz for bulk diamond NV-centers, which is within reach for the mechanical frequencies in our current geometry.
Hence, this system can reach the high cooperativity ($C>1$) and even the strong coupling regime ($\lambda > 2\pi/\tilde{T}_2, \Gamma_\text{th}$). 
Such a strong coupling enables ground-state cooling,
quantum-by-quantum generation of arbitrary states of motion \cite{Rabl:2009fz}, and spin-spin entanglement \cite{Schuetz:2017bl}.

%
Besides the translational degrees of freedom, levitated particles are free to rotate.
For the hard magnets used in our experiment, the anisotropy energy strongly couples the particle orientation to the magnetization axis. The coupling leads to hybrid magneto-rotational modes, which correspond to a librational mode at frequency $\omega_l = \sqrt{\omega_L\omega_I}$ that precesses around the local magnetic field $\mathbf{B}_0$ at the Einstein-deHaas frequency $\omega_I = \densitymag V_\text{mag} / (I_0\gamma_0)$ due to the intrinsic spin angular momentum of the polarized electrons in the magnet.
Here, $\mathbf{B}_0$ is the sum of the field due to the sc and additional external fields and $I_0=2\densitymass V_\text{mag} \a^2/5$ is the moment of inertia, $V_\text{mag}$ being the volume of the magent and $\densitymass$ ($\densitymag$) its mass (spin) density.
Since the Larmor frequency  $\omega_L = \gamma_0 B_0$,  $\omega_l$ is higher than the translational mode frequencies even without additional fields and can be tuned with moderate magnetic fields $\sim 10 \rm mT$ to MHz frequencies.
The high frequencies of the librational modes make them inaccessible to our current detection based on video analysis and DC magnetometry. Future work will explore these modes using optical interferometry and SQUID \cite{PratCamps:2017ci} or NV-AC magnetometry \cite{Kolkowitz:2012iw, Huillery:2019vz}.
The rotational modes couple to the NV-center with a dipole-dipole coupling  $ \lambda_{dp} =\gamma_e \mu_0 m_\text{zp} \frac{a^3}{r'^3} f_{dp}(\theta, \vphi)$, where $f_{dp}(\theta, \vphi)$ is a function $\sim 1$ that depends on the relative position and orientation of the NV-center and the magnet and $m_\text{zp} = \sqrt{\hbar \gamma_e \densitymag \left/2 V_\text{mag}\right.}$ is the zero point magnetization of the Kittel magnon \cite{GonzalezBallestero:2019wi}.
The weaker distance dependence yields $\lambda_{dp} = 0.4 \kHz$ for a $a=5\um$ magnet at $5 \um$ distance from the NV-center  (\figref{fig:Couplings}b).
This parameter regime is readily accessible with the experimental approach presented here, and it is sufficient to probabilistically cool the librational mode near its ground state \cite{Rao:2016iv} (based on the requirements for a harmonic oscillator mode $\omega_0/2\pi\sim \rm MHz$, $\lambda/2\pi\sim 100 \rm Hz$, $Q>10^6$).

\paragraph*{Outlook}
These considerations indicate that our approach is a promising platform for quantum nanomechanics.
Our experimental technique also allows us to achieve levitation with the superconductor in the Meissner state, and thus presents a path forward to observe precession due to the intrinsic spin angular momentum of the magnet with applications in highly sensitive magnetometry \cite{JacksonKimball:2016cq}.
Since our mechanical resonator is all magnetic, we maximize the spin to mass ratio $\densitymag/\sqrt{\densitymass}$ for a given magnetic material, which maximizes the spin-mechanical coupling. This leads to strong spin-mechanical coupling even for moderate experimental parameters.
In addition, this system features libration modes, which are expected to reach an unprecedented spin-mechanical parameter regime even for magnets with $\a\sim 5 \um$. Magnets of this size can be levitated with the experimental technique introduced here. 
The combination of high mechanical Q-factor, strong spin-mechanical coupling, and long spin-coherence is key for a range of applications such as magnetometers, accelerometers and gyroscopes \cite{PratCamps:2017ci}, where the magnet is the sensor which is read out through the NV-center \cite{Kolkowitz:2012iw}. Furthermore, it may enable the exploration of new phenomena, including dynamics between a levitated nanomagnet and a single flux vortex\cite{Thiel:2016es, Pelliccione:2016de, Yechezkel:2018cy}, precession of a non-rotating magnet due to its intrinsic spin angular momentum \cite{Rusconi:2017cm}, preparation of non-Gaussian quantum states \cite{Rabl:2009fz}, 
mechanically mediated quantum networks\cite{Rabl:2010kk, Rusconi:2019bu}, detection of dark matter\cite{JacksonKimball:2016cq}, and measuring the magnets internal degrees of freedom \cite{GonzalezBallestero:2019wi}.\\

This work was supported by the NSF, the Center for Ultracold Atoms (CUA), the ONR MURI Quantum Opto-Mechanics with Atoms and Nanostructured Diamond (QOMAND), the Vannever Bush Faculty Fellowship, and the Moore Foundation.
This material is based upon work supported by the National Science Foundation Graduate Research Fellowship Program under Grant No. DGE1144152 and DGE1745303. Any opinions, findings, and conclusions or recommendations expressed in this material are those of the author(s) and do not necessarily reflect the views of the National Science Foundation.
JG was supported by the European Union (SEQOO, H2020-MSCA-IF-2014, no. 655369).
AK and AS were supported by the Department of Defense (DoD) through the National Defense Science and Engineering Graduate Fellowship (NDSEG) Program. 
C.~G.~B. was supported by the European Union (PWAQUTEC, H2020-MSCA-IF-2017  no.~796725).
This work was performed in part at the Center for Nanoscale Systems (CNS), a member of the National Nanotechnology Coordinated Infrastructure Network (NNCI), which is supported by the National Science Foundation under NSF award no. 1541959. CNS is part of Harvard University. We gratefully acknowledge Frank Zhao for assistance with sample magnetization.

\bibliographystyle{unsrt}

\begin{thebibliography}{10}

\bibitem{Rabl:2010cm}
P~Rabl.
\newblock {Cooling of mechanical motion with a two-level system: The
  high-temperature regime}.
\newblock {\em Physical Review B}, 82(16):165320, October 2010.

\bibitem{Schuetz:2017bl}
M~J~A Schuetz, G~Giedke, L~M~K Vandersypen, and J~I Cirac.
\newblock {High-fidelity hot gates for generic spin-resonator systems}.
\newblock {\em Physical Review A}, 95(5):052335--26, May 2017.

\bibitem{Rabl:2010kk}
P~Rabl, S~J Kolkowitz, F~H~L Koppens, J~G~E Harris, P~Zoller, and Mikhail~D
  Lukin.
\newblock {A quantum spin transducer based on nanoelectromechanical resonator
  arrays}.
\newblock {\em Nature Physics}, 6(8):602--608, May 2010.

\bibitem{Kolkowitz:2012iw}
S~Kolkowitz, A~C Bleszynski~Jayich, Q~P Unterreithmeier, S~D Bennett, P~Rabl,
  J~G~E Harris, and Mikhail~D Lukin.
\newblock {Coherent Sensing of a Mechanical Resonator with a Single-Spin
  Qubit}.
\newblock {\em Science}, 335(6076):1603--1606, March 2012.

\bibitem{Bennett:2013ga}
S~D Bennett, N~Y Yao, J~Otterbach, P~Zoller, P~Rabl, and Mikhail~D Lukin.
\newblock {Phonon-Induced Spin-Spin Interactions in Diamond Nanostructures:
  Application to Spin Squeezing}.
\newblock {\em Physical Review Letters}, 110(15):156402, April 2013.

\bibitem{Barson:2017ba}
Michael S~J Barson, Phani Peddibhotla, Preeti Ovartchaiyapong, Kumaravelu
  Ganesan, Richard~L Taylor, Matthew Gebert, Zoe Mielens, Berndt Koslowski,
  David~A Simpson, Liam~P McGuinness, Jeffrey McCallum, Steven Prawer, Shinobu
  Onoda, Takeshi Ohshima, Ania~C Bleszynski~Jayich, Fedor Jelezko, Neil~B
  Manson, and Marcus~W Doherty.
\newblock {Nanomechanical Sensing Using Spins in Diamond}.
\newblock {\em Nano Letters}, 17(3):1496--1503, March 2017.

\bibitem{vanWezel:2011fy}
J~van Wezel and T~H Oosterkamp.
\newblock {A nanoscale experiment measuring gravity's role in breaking the
  unitarity of quantum dynamics}.
\newblock {\em Proceedings of the Royal Society A: Mathematical, Physical and
  Engineering Sciences}, 468(2137):35--56, November 2011.

\bibitem{Marshall:2003dj}
William Marshall, Christoph Simon, Roger Penrose, and Dik Bouwmeester.
\newblock {Towards quantum superpositions of a mirror}.
\newblock {\em Physical Review Letters}, 91:130401, September 2003.

\bibitem{Rabl:2009fz}
P~Rabl, P~Cappellaro, M~Dutt, L~Jiang, J~Maze, and Mikhail~D Lukin.
\newblock {Strong magnetic coupling between an electronic spin qubit and a
  mechanical resonator}.
\newblock {\em Physical Review B}, 79(4):041302, January 2009.

\bibitem{Arcizet:2011cg}
O~Arcizet, V~Jacques, A~Siria, P~Poncharal, P~Vincent, and S~Seidelin.
\newblock {A single nitrogen-vacancy defect coupled to a nanomechanical
  oscillator}.
\newblock {\em Nature Physics}, 7(11):879--883, September 2011.

\bibitem{Pigeau:2015kl}
B~Pigeau, S~Rohr, L~Mercier de~L~eacute pinay, A~Gloppe, V~Jacques, and
  O~Arcizet.
\newblock {Observation of a phononic Mollow triplet in a multimode hybrid
  spin-nanomechanical system}.
\newblock {\em Nature Communications}, 6:8603, October 2015.

\bibitem{Mintert:2001dy}
Florian Mintert and Christof Wunderlich.
\newblock {Ion-Trap Quantum Logic Using Long-Wavelength Radiation}.
\newblock {\em Physical Review Letters}, 87(25):257904--5, November 2001.

\bibitem{Hong:2012hv}
Sungkun Hong, Michael~S Grinolds, Patrick Maletinsky, Ronald~L Walsworth,
  Mikhail~D Lukin, and Amir Yacoby.
\newblock {Coherent, Mechanical Control of a Single Electronic Spin}.
\newblock {\em Nano Letters}, 12(8):3920--3924, August 2012.

\bibitem{RomeroIsart:2010iv}
Oriol Romero-Isart, Mathieu~L Juan, Romain Quidant, and J~Ignacio Cirac.
\newblock {Toward quantum superposition of living organisms}.
\newblock {\em New Journal of Physics}, 12(3):033015, March 2010.

\bibitem{Chang:2010jn}
Darrick~E Chang, C~A Regal, S~B Papp, D~J Wilson, J~Ye, O~Painter, H~Jeff
  Kimble, and P~Zoller.
\newblock {Cavity opto-mechanics using an optically levitated nanosphere}.
\newblock {\em Proceedings of the National Academy of Sciences},
  107(3):1005--1010, January 2010.

\bibitem{Jain:2016fj}
Vijay Jain, Jan Gieseler, Clemens Moritz, Christoph Dellago, Romain Quidant,
  and Lukas Novotny.
\newblock {Direct Measurement of Photon Recoil from a Levitated Nanoparticle}.
\newblock {\em Physical Review Letters}, 116(24):243601, June 2016.

\bibitem{RomeroIsart:2012hg}
O~Romero-Isart, L~Clemente, C~Navau, A~Sanchez, and J~I Cirac.
\newblock {Quantum Magnetomechanics with Levitating Superconducting
  Microspheres}.
\newblock {\em Physical Review Letters}, 109(14):147205--5, October 2012.

\bibitem{Cirio:2012gka}
M~Cirio, G~K Brennen, and J~Twamley.
\newblock {Quantum Magnetomechanics: Ultrahigh- Q-Levitated Mechanical
  Oscillators}.
\newblock {\em Physical Review Letters}, 109(14):147206--5, October 2012.

\bibitem{Doherty:2013bu}
Marcus~W Doherty, Neil~B Manson, Paul Delaney, Fedor Jelezko, J{\"o}rg
  Wrachtrup, and Lloyd C~L Hollenberg.
\newblock {Physics Reports}.
\newblock {\em Physics Reports}, 528(1):1--45, July 2013.

\bibitem{Taylor:2008cp}
J~M Taylor, P~Cappellaro, L~Childress, L~Jiang, D~Budker, P~R Hemmer, A~Yacoby,
  R~Walsworth, and M~D Lukin.
\newblock {High-sensitivity diamond magnetometer with nanoscale resolution}.
\newblock {\em Nature Physics}, 4(10):810--816, September 2008.

\bibitem{Maze:2008cs}
J~R Maze, P~L Stanwix, J~S Hodges, S~Hong, J~M Taylor, P~Cappellaro, L~Jiang,
  M~V~Gurudev Dutt, E~Togan, A~S Zibrov, A~Yacoby, R~L Walsworth, and M~D
  Lukin.
\newblock {Nanoscale magnetic sensing with an individual electronic spin in
  diamond}.
\newblock {\em Nature}, 455(7213):644--647, October 2008.

\bibitem{Kolkowitz:2015fx}
S~Kolkowitz, A~Safira, A~A High, R~C Devlin, S~Choi, Q~P Unterreithmeier,
  D~Patterson, A~S Zibrov, V~E Manucharyan, H~Park, and M~D Lukin.
\newblock {Probing Johnson noise and ballistic transport in normal metals with
  a single-spin qubit}.
\newblock {\em Science}, 347:1129--1132, March 2015.

\bibitem{Hensen:2015dw}
B~Hensen, H~Bernien, A~E Dr{\'e}au, A~Reiserer, N~Kalb, M~S Blok, J~Ruitenberg,
  R~F~L Vermeulen, R~N Schouten, C~Abell{\'a}n, W~Amaya, V~Pruneri, M~W
  Mitchell, M~Markham, D~J Twitchen, D~Elkouss, S~Wehner, T~H Taminiau, and
  R~Hanson.
\newblock {Loophole-free Bell inequality violation using electron spins
  separated by 1.3 kilometres}.
\newblock {\em Nature}, 526(7575):682--686, October 2015.

\bibitem{Barowski:1993uya}
H~Barowski, K~M Sattler, and W~Schoepe.
\newblock {Static and dynamic forces on a permanent magnet levitating between
  superconducting surfaces}.
\newblock {\em Journal of Low Temperature Physics}, 93(1):85--100, 1993.

\bibitem{Gloos:1990ta}
K~Gloos, J~H KOIVUNIEMI, W~SCHOEPEN, J~T Simola, and J~T TUORINIEMI.
\newblock {VISCOMETER UTILIZING A FLOATING CHARGED MAGNETIC PARTICLE}.
\newblock {\em Physica B: Condensed {\ldots}}, pages 119--120, July 1990.

\bibitem{Druge:2014do}
J~Druge, C~Jean, O~Laurent, M-A M{\'e}asson, and I~Favero.
\newblock {Damping and non-linearity of a levitating magnet in rotation above a
  superconductor}.
\newblock {\em New Journal of Physics}, page 075011, July 2014.

\bibitem{Nemoshkalenko:1990ku}
V~V Nemoshkalenko, E~H Brandt, A~A Kordyuk, and B~G Nikitin.
\newblock {Dynamics of a permanent magnet levitating above a high-Tc
  superconductor}.
\newblock {\em Physica C: Superconductivity}, 170(5-6):481--485, October 1990.

\bibitem{Hull:1999jb}
John~R Hull and Ahmet Cansiz.
\newblock {Vertical and lateral forces between a permanent magnet and a
  high-temperature superconductor}.
\newblock {\em Journal of Applied Physics}, 86(11):6396--6404, December 1999.

\bibitem{Tao:2019kl}
Tao Wang, Sean Lourette, Sean~R O{\textquoteright}Kelley, Metin Kayci, Y~B
  Band, Derek F~Jackson Kimball, Alexander~O Sushkov, and Dmitry Budker.
\newblock {Dynamics of a Ferromagnetic Particle Levitated over a
  Superconductor}.
\newblock {\em Physical Review Applied}, 11(4):044041, April 2019.

\bibitem{Neukirch:2013ua}
Levi~P Neukirch, Jan Gieseler, Romain Quidant, Lukas Novotny, and
  A~Nick~Vamivakas.
\newblock {Observation of nitrogen vacancy photoluminescence from an optically
  levitated nanodiamond}.
\newblock {\em Optics letters}, 38(16):2976--2979, January 2013.

\bibitem{Hoang:2016ky}
Thai~M Hoang, Jonghoon Ahn, Jaehoon Bang, and Tongcang Li.
\newblock {Electron spin control of optically levitated nanodiamonds in
  vacuum}.
\newblock {\em Nature Communications}, 7:1--8, July 2016.

\bibitem{Kuhlicke:2014dm}
Alexander Kuhlicke, Andreas~W Schell, Joachim Zoll, and Oliver Benson.
\newblock {Nitrogen vacancy center fluorescence from a submicron diamond
  cluster levitated in a linear quadrupole ion trap}.
\newblock {\em Applied Physics Letters}, 105(7):073101, August 2014.

\bibitem{Delord:2017cb}
T~Delord, L~Nicolas, L~Schwab, and G~H{\'e}tet.
\newblock {Electron spin resonance from NV centers in diamonds levitating in an
  ion trap}.
\newblock {\em New Journal of Physics}, 19(3):033031--11, March 2017.

\bibitem{Alda:2016ce}
I~Alda, J~Berthelot, R~A Rica, and Romain Quidant.
\newblock {Trapping and manipulation of individual nanoparticles in a planar
  Paul trap}.
\newblock {\em Applied Physics Letters}, 109(16):163105--5, October 2016.

\bibitem{Hsu:2016cs}
Jen-Feng Hsu, Peng Ji, Charles~W Lewandowski, and Brian D{\textquoteright}Urso.
\newblock {Cooling the Motion of Diamond Nanocrystals in a Magneto-
  Gravitational Trap in High Vacuum}.
\newblock {\em Scientific Reports}, 6:30125, July 2016.

\bibitem{OBrien:2019kc}
M~C O'Brien, S~Dunn, J~E Downes, and J~Twamley.
\newblock {Magneto-mechanical trapping of micro-diamonds at low pressures}.
\newblock {\em Applied Physics Letters}, 114(5):053103--6, February 2019.

\bibitem{Gieseler:2012bi}
Jan Gieseler, B~Deutsch, Romain Quidant, and Lukas Novotny.
\newblock {Subkelvin Parametric Feedback Cooling of a Laser-Trapped
  Nanoparticle}.
\newblock {\em Physical Review Letters}, 109(10):103603, 2012.

\bibitem{Huillery:2019vz}
P~Huillery, T~Delord, L~Nicolas, M~Van Den~Bossche, M~Perdriat, and G.~Hetet.
\newblock {Spin-mechanics with levitating ferromagnetic particles}.
\newblock {\em arXiv.org}.

\bibitem{Skomski:2003by}
R~Skomski.
\newblock {Nanomagnetics}.
\newblock {\em Journal of Physics: Condensed Matter}, 15(20):R841--R896, May
  2003.

\bibitem{Norte:2016eq}
R~A Norte, J~P Moura, and S~Gr{\"o}blacher.
\newblock {Mechanical Resonators for Quantum Optomechanics Experiments at Room
  Temperature}.
\newblock {\em Physical Review Letters}, 116(14):021001--6, April 2016.

\bibitem{Reinhardt:2016ef}
Christoph Reinhardt, Tina M{\"u}ller, Alexandre Bourassa, and Jack~C Sankey.
\newblock {Ultralow-Noise SiN Trampoline Resonators for Sensing and
  Optomechanics}.
\newblock {\em Physical Review X}, 6(2):021001--8, April 2016.

\bibitem{Tsaturyan:2017kd}
Y~Tsaturyan, A~Barg, E~S Polzik, and A~Schliesser.
\newblock {Ultracoherent nanomechanical resonators via soft clamping and
  dissipation dilution}.
\newblock 12(8):776--783, June 2017.

\bibitem{Ghadimi:2018kb}
A~H Ghadimi, S~A Fedorov, N~J Engelsen, M~J Bereyhi, R~Schilling, D~J Wilson,
  and Tobias~J Kippenberg.
\newblock {Elastic strain engineering for ultralow mechanical dissipation}.
\newblock {\em Science}, 360(6390):764--768, May 2018.

\bibitem{Longenecker:2012ip}
Jonilyn~G Longenecker, H~J Mamin, Alexander~W Senko, Lei Chen, Charles~T
  Rettner, Daniel Rugar, and John~A Marohn.
\newblock {High-Gradient Nanomagnets on Cantilevers for Sensitive Detection of
  Nuclear Magnetic Resonance}.
\newblock {\em ACS Nano}, 6(11):9637--9645, November 2012.

\bibitem{Fischer:2019fy}
R~Fischer, D~P McNally, C~Reetz, G~G~T Assump{\c c}{\~a}o, T~Knief, Y~Lin, and
  C~A Regal.
\newblock {Spin detection with a micromechanical trampoline: towards magnetic
  resonance microscopy harnessing cavity optomechanics}.
\newblock {\em New Journal of Physics}, 21(4):043049--14, April 2019.

\bibitem{Tetienne:2012fv}
J-P Tetienne, L~Rondin, P~Spinicelli, M~Chipaux, T~Debuisschert, J-F Roch, and
  V~Jacques.
\newblock {Magnetic-field-dependent photodynamics of single NV defects in
  diamond: an application to qualitative all-optical magnetic imaging}.
\newblock {\em New Journal of Physics}, 14(10):103033--16, October 2012.

\bibitem{Rondin:2014kd}
L~Rondin, J-P Tetienne, T~Hingant, J-F Roch, P~Maletinsky, and V~Jacques.
\newblock {Magnetometry with nitrogen-vacancy defects in diamond}.
\newblock {\em Reports on Progress in Physics}, 77(5):056503, May 2014.

\bibitem{FornDiaz:2019br}
P~Forn-D{\'\i}az, L~Lamata, E~Rico, J~Kono, and E~Solano.
\newblock {Ultrastrong coupling regimes of light-matter interaction}.
\newblock {\em Reviews of Modern Physics}, 91(2):025005, June 2019.

\bibitem{Gieseler:2013ip}
Jan Gieseler, Lukas Novotny, and Romain Quidant.
\newblock {Thermal nonlinearities in a nanomechanical oscillator}.
\newblock {\em Nature Physics}, 9(12):806--810, November 2013.

\bibitem{Abobeih:2018bc}
M~H Abobeih, J~Cramer, M~A Bakker, N~Kalb, M~Markham, D~J Twitchen, and T~H
  Taminiau.
\newblock {One-second coherence for a single electron spin coupled to a
  multi-qubit nuclear-spin environment}.
\newblock {\em Nature Communications}, 9:2552, June 2018.

\bibitem{BarGill:2013dq}
N~Bar-Gill, L~M Pham, A~Jarmola, D~Budker, and R~L Walsworth.
\newblock {Solid-state electronic spin coherence time approaching one second}.
\newblock {\em Nature Communications}, 4:1743--6, April 2013.

\bibitem{Kordyuk:1998ht}
Alexander~A Kordyuk.
\newblock {Magnetic levitation for hard superconductors}.
\newblock {\em Journal of Applied Physics}, 83(1):610--612, June 1998.

\bibitem{PratCamps:2017ci}
J~Prat-Camps, C~Teo, C~C Rusconi, W~Wieczorek, and O~Romero-Isart.
\newblock {Ultrasensitive Inertial and Force Sensors with Diamagnetically
  Levitated Magnets}.
\newblock {\em Physical Review Applied}, 8(3):034002, September 2017.

\bibitem{GonzalezBallestero:2019wi}
Carlos Gonzalez-Ballestero, Jan Gieseler, and Oriol Romero-Isart.
\newblock {Quantum Acoustomechanics with a Micromagnet}.
\newblock {\em arXiv.org}, July 2019.

\bibitem{Rao:2016iv}
D~D~Bhaktavatsala Rao, S~Ali Momenzadeh, and J{\"o}rg Wrachtrup.
\newblock {Heralded Control of Mechanical Motion by Single Spins}.
\newblock {\em Physical Review Letters}, 117(7):077203--8, August 2016.

\bibitem{JacksonKimball:2016cq}
Derek~F Jackson~Kimball, Alexander~O Sushkov, and Dmitry Budker.
\newblock {Precessing Ferromagnetic Needle Magnetometer}.
\newblock {\em Physical Review Letters}, 116(19):190801, May 2016.

\bibitem{Thiel:2016es}
L~Thiel, D~Rohner, M~Ganzhorn, P~Appel, E~Neu, B~M{\"u}ller, R~Kleiner,
  D~Koelle, and P~Maletinsky.
\newblock {Quantitative nanoscale vortex imaging using a cryogenic quantum
  magnetometer}.
\newblock 11(8):677 --681, May 2016.

\bibitem{Pelliccione:2016de}
Matthew Pelliccione, Alec Jenkins, Preeti Ovartchaiyapong, Christopher Reetz,
  Eve Emmanouilidou, Ni~Ni, and Ania C~Bleszynski Jayich.
\newblock {Scanned probe imaging of nanoscale magnetism at cryogenic
  temperatures with a single-spin quantum sensor}.
\newblock {\em Nature Nanotechnology}, 11(8):700 --705, May 2016.

\bibitem{Yechezkel:2018cy}
Schlussel Yechezkel, Lenz Till, Rohner Dominik, Bar-Haim Yaniv, Bougas
  Lykourgos, Kieschnick Michael, Rozenberg Evgeny, Thiel Lucas, Waxman Amir,
  Maletinsky Patrick, Budker Dmitry, and Folman Ron.
\newblock {Wide-Field Imaging of Superconductor Vortices with Electron Spins in
  Diamond}.
\newblock {\em Physical Review Applied}, 10(1):1, September 2018.

\bibitem{Rusconi:2017cm}
C~C Rusconi, V~P{\"o}chhacker, K~Kustura, J~I Cirac, and O~Romero-Isart.
\newblock {Quantum Spin Stabilized Magnetic Levitation}.
\newblock {\em Physical Review Letters}, 119(16):167202, October 2017.

\bibitem{Rusconi:2019bu}
C~C Rusconi, M~J~A Schuetz, Jan Gieseler, M~D Lukin, and O~Romero-Isart.
\newblock {Hybrid architecture for engineering magnonic quantum networks}.
\newblock {\em Physical Review A}, 100(2):022343, August 2019.
\end{thebibliography}

\end{document}